\documentclass[twocolumn,usenatbib]{aastex62}

\usepackage{graphicx}	
\usepackage{amsmath}	
\usepackage{amssymb}	

\received{\today}
\revised{\today}
\accepted{\today}
\submitjournal{ApJ}

\shorttitle{Radiative electron-ion turbulence}
\shortauthors{Zhdankin et al.}

\begin{document}

\newcommand{\red}{\textcolor{red}}
\newcommand{\blue}{\textcolor{blue}}
\newcommand{\green}{\textcolor{green}}

\title{Production and persistence of extreme two-temperature plasmas in radiative relativistic turbulence} 

\correspondingauthor{Vladimir Zhdankin}
\email{zhdankin@princeton.edu}

\author{Vladimir Zhdankin}
\thanks{NASA Einstein fellow}
\affiliation{Department of Astrophysical Sciences, Princeton University, 4 Ivy Lane, Princeton, NJ 08544, USA}

\author{Dmitri A. Uzdensky}
\affiliation{Center for Integrated Plasma Studies, Department of Physics, 390 UCB, University of Colorado, Boulder, CO 80309, USA}

\author{Matthew W. Kunz}
\affiliation{Department of Astrophysical Sciences, Princeton University, 4 Ivy Lane, Princeton, NJ 08544, USA} 
\affiliation{Princeton Plasma Physics Laboratory, P.O. Box 451, Princeton, NJ 08543, USA}




\begin{abstract} 
Turbulence is a predominant process for energizing electrons and ions in collisionless astrophysical plasmas, and thus is responsible for shaping their radiative signatures (luminosity, spectra, and variability). To better understand the kinetic properties of a collisionless radiative plasma subject to externally driven turbulence, we investigate particle-in-cell simulations of relativistic plasma turbulence with external inverse Compton cooling acting on the electrons. We find that ions continuously heat up while electrons gradually cool down (due to the net effect of radiation), and hence the ion-to-electron temperature ratio $T_i/T_e$ grows in time. We show that $T_i/T_e$ is limited only by the size and duration of the simulations (reaching $T_i/T_e \sim 10^3$), indicating that there are no efficient collisionless mechanisms of electron-ion thermal coupling. This result has implications for models of radiatively inefficient accretion flows, such as observed in the Galactic Center and in M87, for which so-called two-temperature plasmas with $T_i/T_e \gg 1$ have been invoked to explain their low luminosity. Additionally, we find that electrons acquire a quasi-thermal distribution (dictated by the competition of turbulent particle energization and radiative cooling), while ions undergo efficient nonthermal acceleration (acquiring a harder distribution than in equivalent non-radiative simulations). There is a modest nonthermal population of high-energy electrons that are beamed intermittently in space, time, and direction; these beamed electrons may explain rapid flares in certain high-energy astrophysical systems (e.g., in the Galactic Center). These numerical results demonstrate that extreme two-temperature plasmas can be produced and maintained by relativistic radiative turbulence.
\end{abstract}



\section{Introduction} \label{sec:intro}

Due to their low densities and extremely high temperatures, many high-energy astrophysical plasmas are collisionless, relativistic, and radiative. Collisionless plasmas are easily perturbed out of thermal equilibrium by turbulent electromagnetic fields. Recent first-principles numerical studies indicate that, in a relativistic plasma, the ensuing nonlinear dynamics lead to rich kinetic phenomena including nonthermal particle acceleration \citep{zhdankin_etal_2017,zhdankin_etal_2018b, comisso_sironi_2018, comisso_sironi_2019, nattila_2019, wong_etal_2020}, the formation of ``two-temperature'' plasmas [\cite{zhdankin_etal_2019}; see also \cite{kawazura_etal_2019}, \cite{arzamasskiy_etal_2019}, and \cite{alves_etal_2019}], and coherent beaming of high-energy particles and photons \citep{zhdankin_etal_2020}. Since these kinetic effects manifest in the spectrum, luminosity, and variability of radiation emitted by the plasma, they have profound implications for astronomical observations.

Kinetic turbulence provides an attractive paradigm for explaining the observed features of active galactic nuclei \citep{yuan_narayan_2014} and black-hole X-ray binaries \citep{remillard_mcclintock_2006}, where plasma surrounding a black hole is collected through a turbulent accretion flow. For example, theoretical models have invoked two-temperature plasmas, where ions become much hotter than electrons due to the unequal deposition of turbulent energy \citep[e.g.,][]{quataert_gruzinov_1999, howes_2010}, to explain the radiative inefficiency of certain classes of these accretion flows \citep{shapiro_lightman_eardley_1976, ichimaru_1977, rees_etal_1982, narayan_yi_1995}. Nonthermal particle acceleration may explain the broadband radiation spectra in these systems, while localized beaming by turbulent structures is a potential mechanism for producing intermittent high-energy flares. 

Radiative cooling (e.g., from synchrotron or inverse Compton processes) may compete with turbulent energization by thermalizing the plasma and maintaining an equilibrium temperature. It has been suggested that radiative cooling may reduce nonthermal electron populations either by steepening the distribution \citep{kardashev_1962, sobacchi_lyubarsky_2020}, by imposing a high-energy cutoff, or by thermalizing the distribution altogether \citep{schlickeiser_1985, zhdankin_etal_2020}. Since ions are typically unaffected by radiative cooling, the ion-to-electron temperature ratio, $T_i/T_e$, will increase over time {\it unless} there exists a sufficiently strong collisionless thermal coupling mechanism between the two species, in which case thermal energy transfer from ions to electrons will limit $T_i/T_e$. Previous studies have proposed mechanisms for such coupling, e.g., unstable modes in small-scale magnetohydrodynamic (MHD) turbulence \citep{begelman_chiueh_1988} and the ion-cyclotron instability \citep{sironi_narayan_2015, sironi_2015}. But whether any such mechanisms can efficiently operate in a turbulent medium remains an open question.

This Letter demonstrates the ability of kinetic turbulence to produce and maintain a two-temperature plasma in the presence of radiative cooling. We apply particle-in-cell (PIC) simulations to investigate the kinetic consequences of driven turbulence in a relativistic electron-ion plasma with strong electron cooling by external inverse Compton (IC) radiation. We show that the plasma acquires a mixture of thermal and nonthermal features: ions are efficiently heated and accelerated, while electrons are predominantly cooled and thermalized, although a modest nonthermal population of intermittently beamed electrons also exists. As a result, an extreme two-temperature plasma with very low radiative efficiency is established.

\section{Methods} \label{sec:methods}

We focus on a relativistically hot plasma, $T_s/m_s c^2 \gg 1$, where $T_s$ and $m_s$ are the temperature and rest mass, respectively, of the electrons ($s = e$) or ions ($s = i$); future work will consider the regime relevant to accretion flows, in which ions are modestly sub-relativistic. In the ultra-relativistic limit, the particle rest masses are negligible compared to their thermal energies. As a result, when $T_i/T_e = 1$, the system behaves as though it were composed of electron-positron pairs. In our setup, this symmetry between the particle species is broken by IC radiation, which acts on electrons but not on ions, causing $T_i/T_e$ (and the electron-ion kinetic scale separation) to grow.

We perform simulations of driven turbulence using the PIC code {\sc Zeltron} \citep{cerutti_etal_2013}, closely following the numerical setup described in \cite{zhdankin_etal_2018}. The domain is a periodic cubic box of volume $L^3$ with mean magnetic field $\boldsymbol{B}_0 = B_0 \hat{\boldsymbol{z}}$. We initialize electrons and ions (protons) from a Maxwell--J\"{u}ttner distribution with number density per species $n_0$ and initial ion temperature $T_{i0} = 100 m_i c^2$ (chosen arbitrarily). We drive strong ($\delta B_{\rm rms} \sim B_0$) turbulence at low wavenumbers ($k \sim 2\pi/L$) by applying a randomly fluctuating external current density \citep{tenbarge_etal_2014}. We incorporate IC cooling from a uniform, time-independent, and isotropic bath of external photons by including a radiation reaction force acting on electrons, $\boldsymbol{F}_{\rm IC} = - (4/3) \sigma_{\rm T} U_{\rm ph} \gamma^2 \boldsymbol{v}/c$, where $\sigma_{\rm T}$ is the Thomson cross section, $U_{\rm ph}$ is the photon energy density, $\boldsymbol{v}$ is the particle velocity, and $\gamma = (1-v^2/c^2)^{-1/2}$ \citep{landau_lifshitz_1975}. We assume an optically thin plasma, so emitted photons escape the system and are not tracked.

A fundamental quantity in this system is the radiative efficiency $\eta_{\rm rad}$, defined as the ratio of the radiative cooling rate $\dot{\epsilon}_{\rm rad}$ to the external energy injection rate $\dot{\epsilon}_{\rm inj}$ (statistically constant in time). These rates are given by
\begin{align}
\dot{\epsilon}_{\rm rad}(t) &= \frac{4}{3} \sigma_{\rm T} c U_{\rm ph} \overline{\gamma_e^2}(t) \, , \nonumber \\
\dot{\epsilon}_{\rm inj} &\sim \frac{B_0^2}{8 \pi n_0} \frac{v_{{\rm A}0}}{L} \, ,
\end{align} 
where $\overline{\gamma_e^2}$ is the mean squared Lorentz factor [$12 (T_e/m_e c^2)^2$ for a relativistic Maxwell--J\"{u}ttner distribution] and $v_{{\rm A}0} = [\sigma_0/(1+\sigma_0)]^{1/2} c$ is the initial Alfv\'{e}n velocity (the initial magnetization $\sigma_0 = 1/2 \beta_0$ in the relativistically hot limit, where plasma $\beta$ is defined below). The radiative efficiency thus scales as
\begin{align}
\eta_{\rm rad} \equiv \frac{\dot{\epsilon}_{\rm rad}}{\dot{\epsilon}_{\rm inj}} &\sim 16 \tau_{\rm T} \frac{8 \pi U_{\rm ph}}{B_0^2} \frac{c}{v_{{\rm A}0}} \left (\frac{T_e}{m_e c^2}\right)^2 \nonumber \\
&\sim \eta_{{\rm rad},0}  \left( \frac{T_e}{T_{e0}} \right)^{2} \, . \label{eq:est}
\end{align}
where $\tau_{\rm T} = \sigma_{\rm T} L n_0$, $T_{e0}$ is the initial electron temperature, and $\eta_{{\rm rad}, 0} \equiv 16 \tau_{\rm T} (8 \pi U_{\rm ph}/B_0^2) (c/v_{{\rm A}0}) (T_{e0}/m_e c^2)^2$ is the initial characteristic radiative efficiency.

The evolution of $T_e$ (and $\eta_{\rm rad}$) is closely tied to the ion-to-electron heating ratio, $Q_i/Q_e$, in a turbulent collisionless plasma. The dependence of $Q_i/Q_e$ on plasma parameters such as $\beta$ and $T_i/T_e$ remains under debate \citep[e.g.,][]{quataert_1998, gruzinov_1998, quataert_gruzinov_1999, howes_2010, zhdankin_etal_2019, kawazura_etal_2019, schekochihin_etal_2019}. We note that, if $Q_i/Q_e$ is a function of the ion plasma beta $\beta_i$ and $T_i/T_e$, and if energy is injected at a constant rate $\dot{\epsilon}_{\rm inj} = Q_i + Q_e$ (with turbulent magnetic energy constant), then there is no IC radiative steady-state value of $T_e$ unless $Q_i/Q_e$ is constant or a function only of $\beta_e = \beta_i T_e/T_i$. This stringent condition arises because, if $Q_i/Q_e$ increases as ions heat up, the electrons would receive a steadily diminishing fraction of the injected energy, and thus their temperature must decrease to maintain an instantaneous equilibrium with radiative cooling.

We characterize the simulations by the following physical parameters (a subscript 0 indicates the initial value): the initial radiative efficiency $\eta_{{\rm rad},0}$; the plasma beta~$\beta \equiv \beta_i + \beta_e$, where~$\beta_s \equiv 8 \pi n_0 T_s/B_{\rm rms}^2$ and~$B_{\rm rms}^2 = B_0^2 + \delta B_{\rm rms}^2$ is the (instantaneous) mean squared magnetic field; $T_i/T_e$; and the driving scale $L/2\pi$ relative to the characteristic ion gyroradius $\rho_i = 3 T_i / e B_{\rm rms}$ (assuming the relativistic limit). Because temperature is ill-defined for a nonthermal plasma, in these definitions we assume $T_s = \overline{\epsilon}_{{\rm kin},s}/3$, where $\overline{\epsilon}_{{\rm kin},s}$ is the average particle kinetic energy (including contributions from bulk motions, which we find to be at most comparable to internal energy).

We primarily focus on results from a fiducial run with physical parameters $\beta_0 = 0.25$ (yielding $\sigma_0 = 2$, $v_{{\rm A}0}/c = 0.82$), $T_{i0}/T_{e0} = 1$, $L/2\pi\rho_{i0} = 40.7$, $\eta_{{\rm rad},0} = 0.9$, and duration $14 L/v_{{\rm A}0}$. The numerical parameters are $N^3 = 1024^3$ cells and cell size $\Delta x = \rho_{i0}/4 = \rho_{e0}/4$. For comparison, we also ran an identical simulation except without radiative cooling ($\eta_{{\rm rad},0} = 0$). We also describe an extreme case, which is similar to the late stages of the fiducial simulation but with coarser resolution, allowing higher $T_i/T_e$. This case has $T_{i0}/T_{e0} = 80$, $\beta_0 = 4$, $L/2\pi\rho_{i0} = 1.5$, $\eta_{{\rm rad},0} = 0.22$, $768^3$ cells, and $\Delta x = \rho_{e0}$; the relativistic electron skin depth is resolved by $d_e = (3 T_e/4\pi n_0 e^2)^{1/2} \approx 2 \Delta x$ at late times (convergence studies confirm that our quantitative results are not affected by numerical resolution). Finally, we describe a parameter scan across $\eta_{{\rm rad},0}$ (changing $U_{\rm ph}$), from simulations with $\beta_0 = 1$, $T_{i0}/T_{e0} = 1$, $L/2\pi\rho_{i0} = 15.3$, $384^3$ cells, $\Delta x = \rho_{e0}/4$, and varying $\eta_{{\rm rad},0} \in \{ 0.028, 0.056, 0.11, 0.22, 0.45, 0.90, 1.8, 3.6\}$. We conducted a broad parameter exploration with additional smaller simulations, which yielded results qualitatively similar to the fiducial run. All simulations use $32$ particles per cell.

\section{Results} \label{sec:results}

\subsection{Spatial structure of temperatures}

\begin{figure}
     \includegraphics[width=\columnwidth]{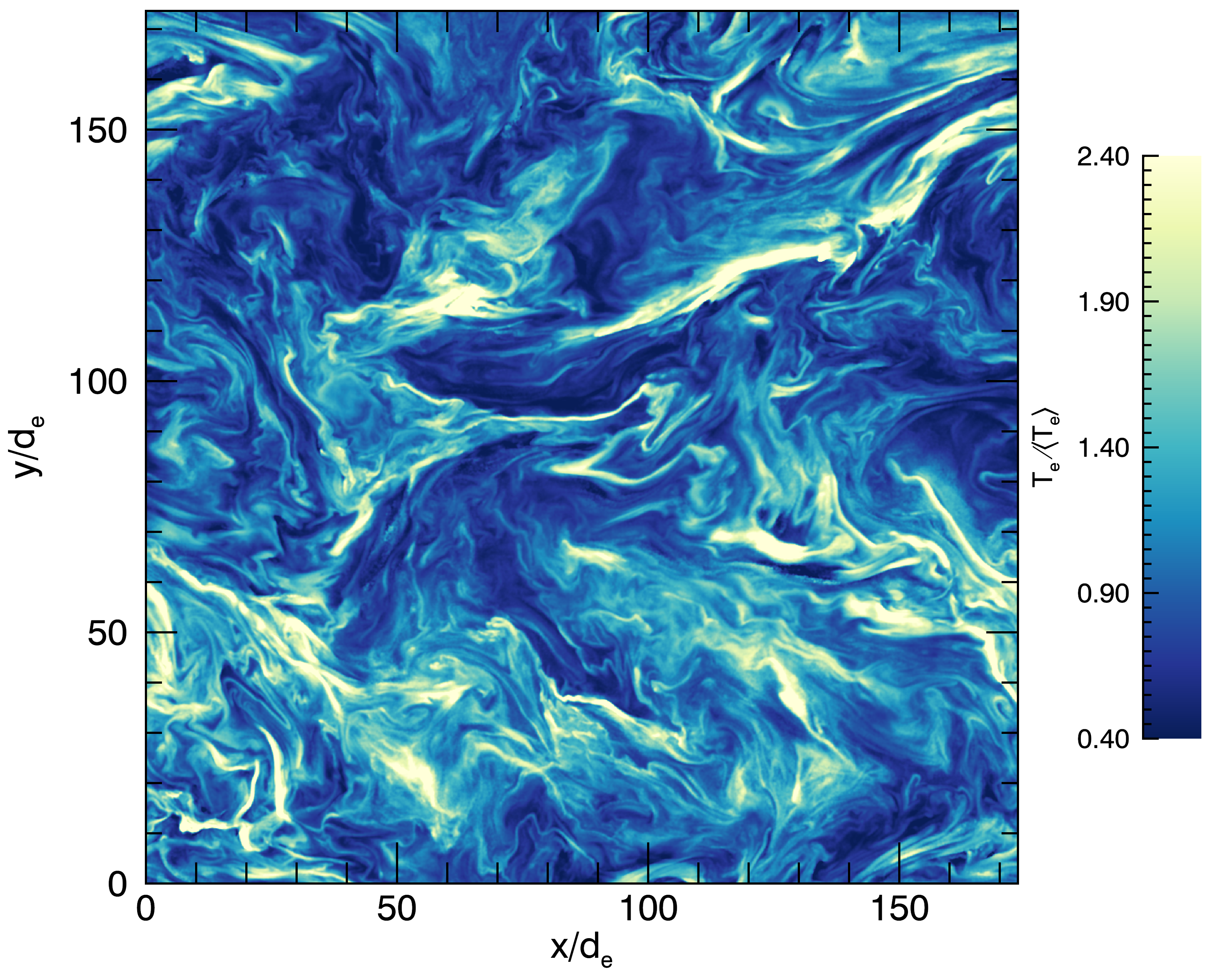}
   \includegraphics[width=\columnwidth]{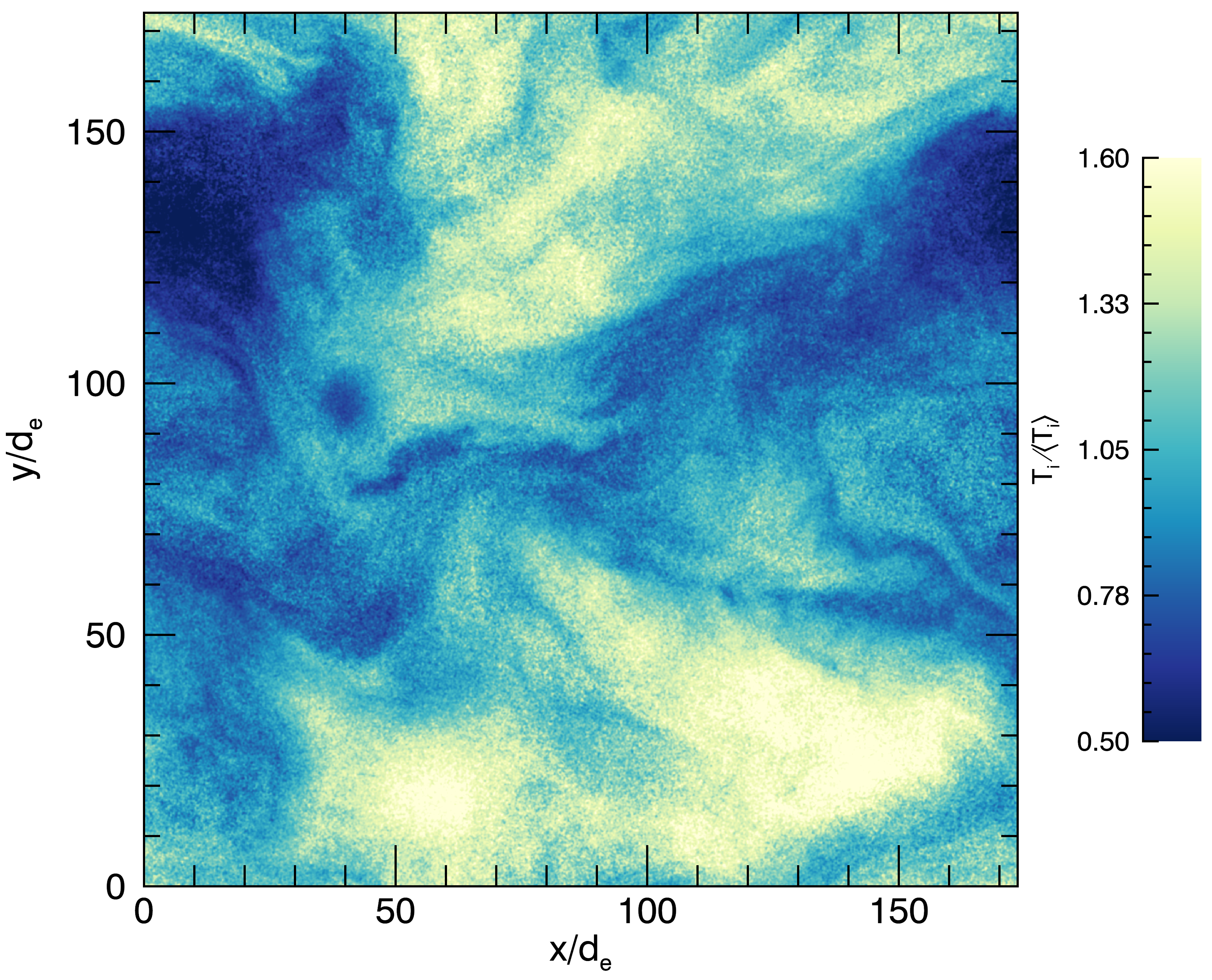}
   \includegraphics[width=\columnwidth]{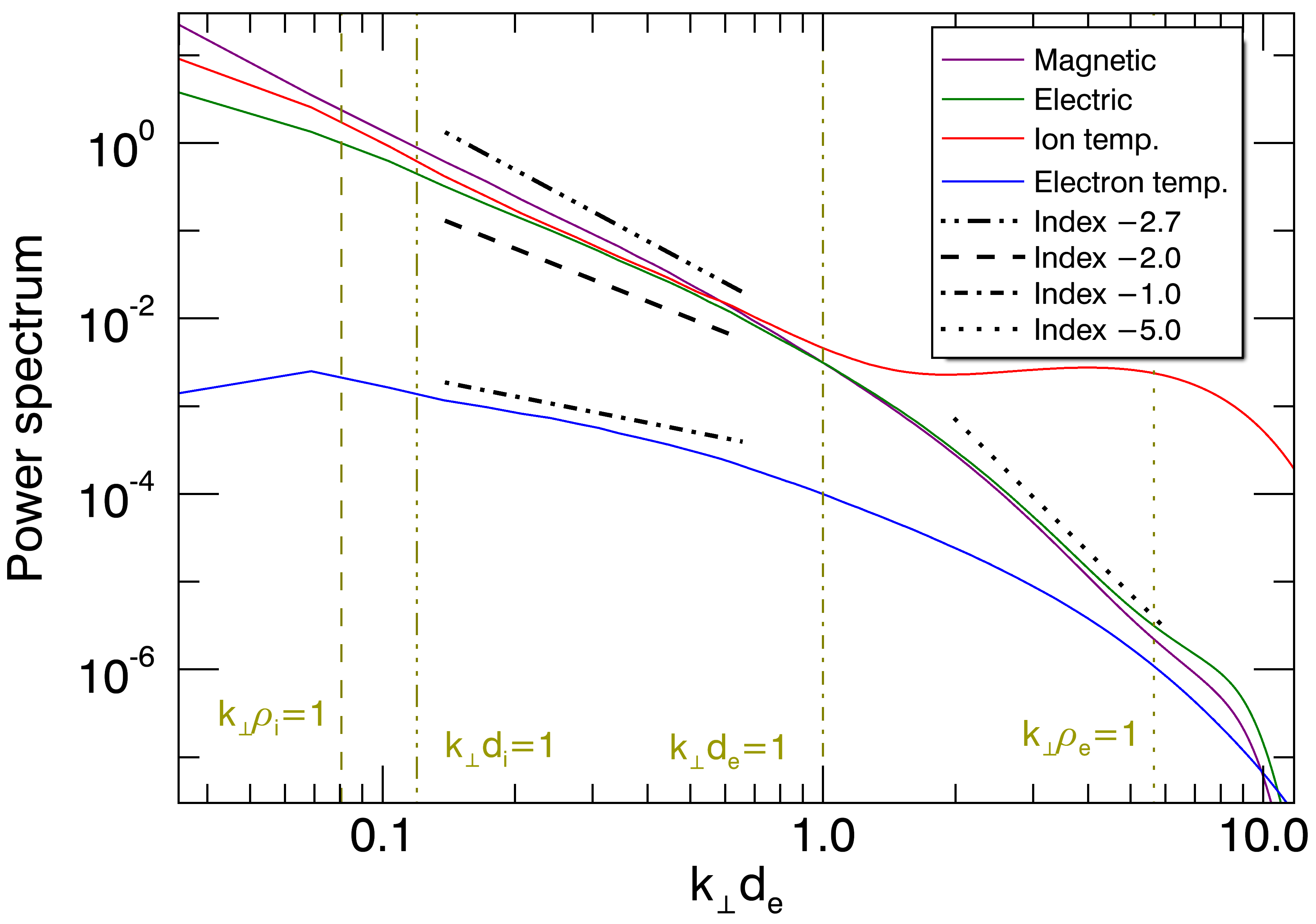}
   \centering
   \caption{\label{fig:images} Top: electron temperature $T_e$ in an $x$-$y$ plane of the fiducial simulation. Middle: same for ion temperature $T_i$. Bottom: power spectra for turbulent magnetic field (purple), electric field (green), $T_i$ (red), and $T_e$ (blue). For reference, power-law scalings (black) and characteristic scales (gold) are indicated.}
 \end{figure}

In Fig.~\ref{fig:images}, we show the electron temperature $T_e$ (top panel) and ion temperature $T_i$ (middle panel) in an arbitrary $x$-$y$ plane from the fiducial simulation at time $t v_{{\rm A}0}/L = 5.5$, when a large ion-to-electron temperature ratio $T_i/T_e \approx 50$ has developed (see Sec.~\ref{sec:heating}). Hot electrons are localized in thin structures with thicknesses near the electron skin depth $d_e$, while hot ions are concentrated in much larger structures at scales $\rho_i \sim d_i$. To characterize the fluctuations, in the bottom panel of Fig.~\ref{fig:images} we show Fourier power spectra (with respect to wavenumber perpendicular to $\boldsymbol{B}_0$, denoted $k_\perp$) for the magnetic field $\boldsymbol{B}$, electric field $\boldsymbol{E}$, $T_i$, and $T_e$; to fit all spectra on the same axes, the latter two are renormalized arbitrarily. These spectra are averaged from $t v_{{\rm A}0}/L = 5.7$ to $t v_{{\rm A}0}/L = 8.0$, during which period the large-scale MHD inertial range is very limited because turbulent heating has caused $\rho_i$ to become comparable to $L/2\pi$. Nevertheless, all of these spectra exhibit power laws between $k_\perp \rho_i = 1$ and $k_\perp d_e = 1$. The power-law index for the magnetic energy spectrum is close to the typical value of $\alpha_B \approx -2.7$ characteristic of a non-relativistic kinetic-Alfv\'{e}n-wave cascade \cite[e.g.,][]{boldyrev_perez_2012}. The spectrum for $T_i$ is similar to the magnetic energy spectrum, while $T_e$ has a much shallower power law with index near $-1$. Note that the $T_i$ spectrum at $k_\perp d_e > 1$ is affected by numerical noise.

\subsection{Turbulent heating and radiative efficiency} \label{sec:heating}

\begin{figure}
    \includegraphics[width=\columnwidth]{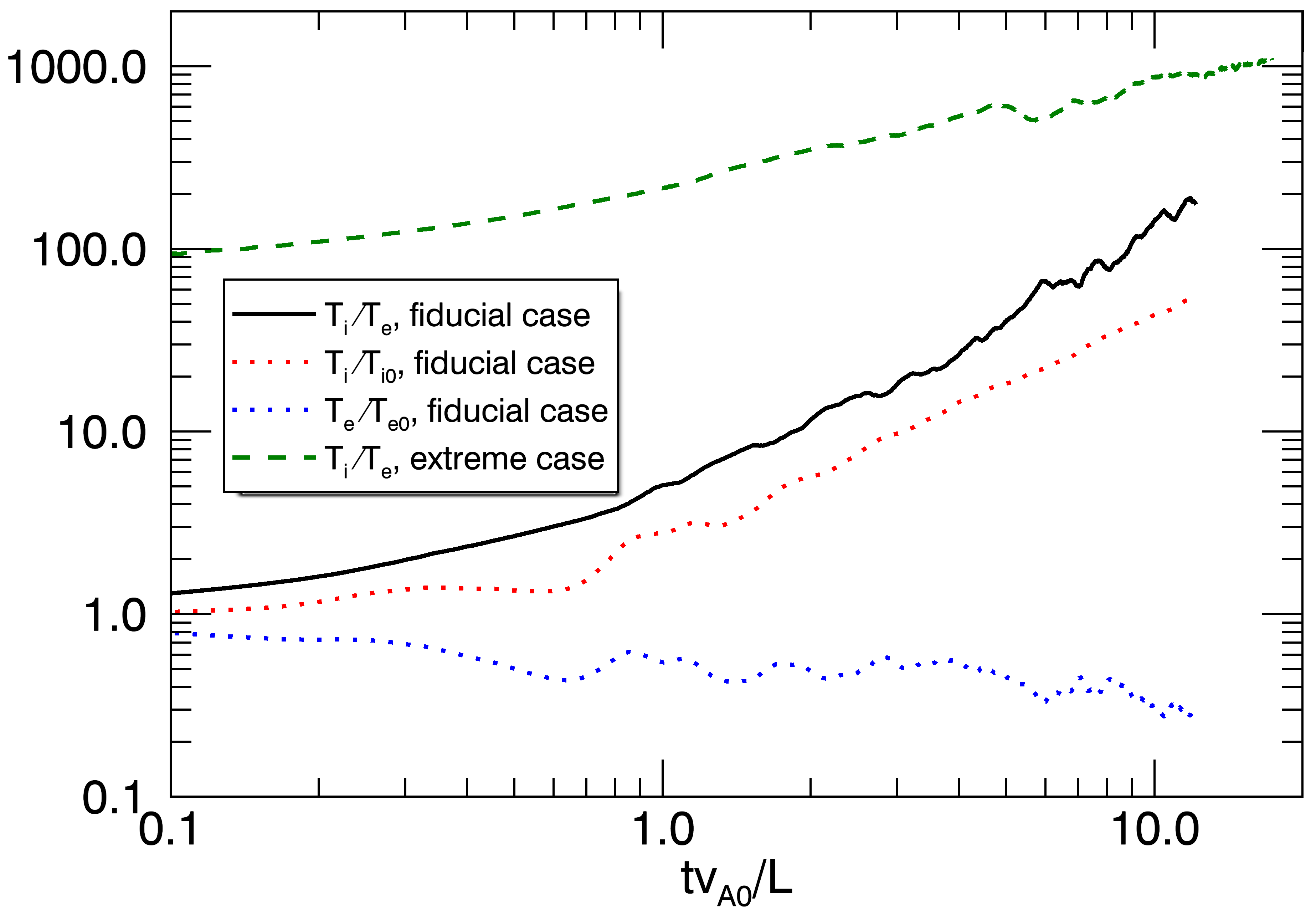}
   \centering
   \caption{\label{fig:params} Evolution of temperature ratio $T_i/T_e$ (black), ion temperature $T_i/T_{i0}$ (red), and electron temperature $T_e/T_{e0}$ (blue) in the fiducial simulation. For comparison, $T_i/T_e$ from the extreme case with $T_{i0}/T_{e0} = 80$ is also shown (green).}
 \end{figure}

As shown in Fig.~\ref{fig:params}, after turbulence fully develops, $T_i$ increases at a roughly constant rate due to turbulent ion heating, while $T_e$ slowly decreases from its initial value due to radiative cooling, which outpaces the turbulent electron heating. Consequently, the temperature ratio increases to $T_i/T_e \gtrsim 10^2$. We also overlay $T_i/T_e$ from the extreme case (green dashed line in Fig.~\ref{fig:params}), which extends the evolution and reaches $T_i/T_e \sim 10^3$ with no indication of saturation.

\begin{figure}
  \includegraphics[width=\columnwidth]{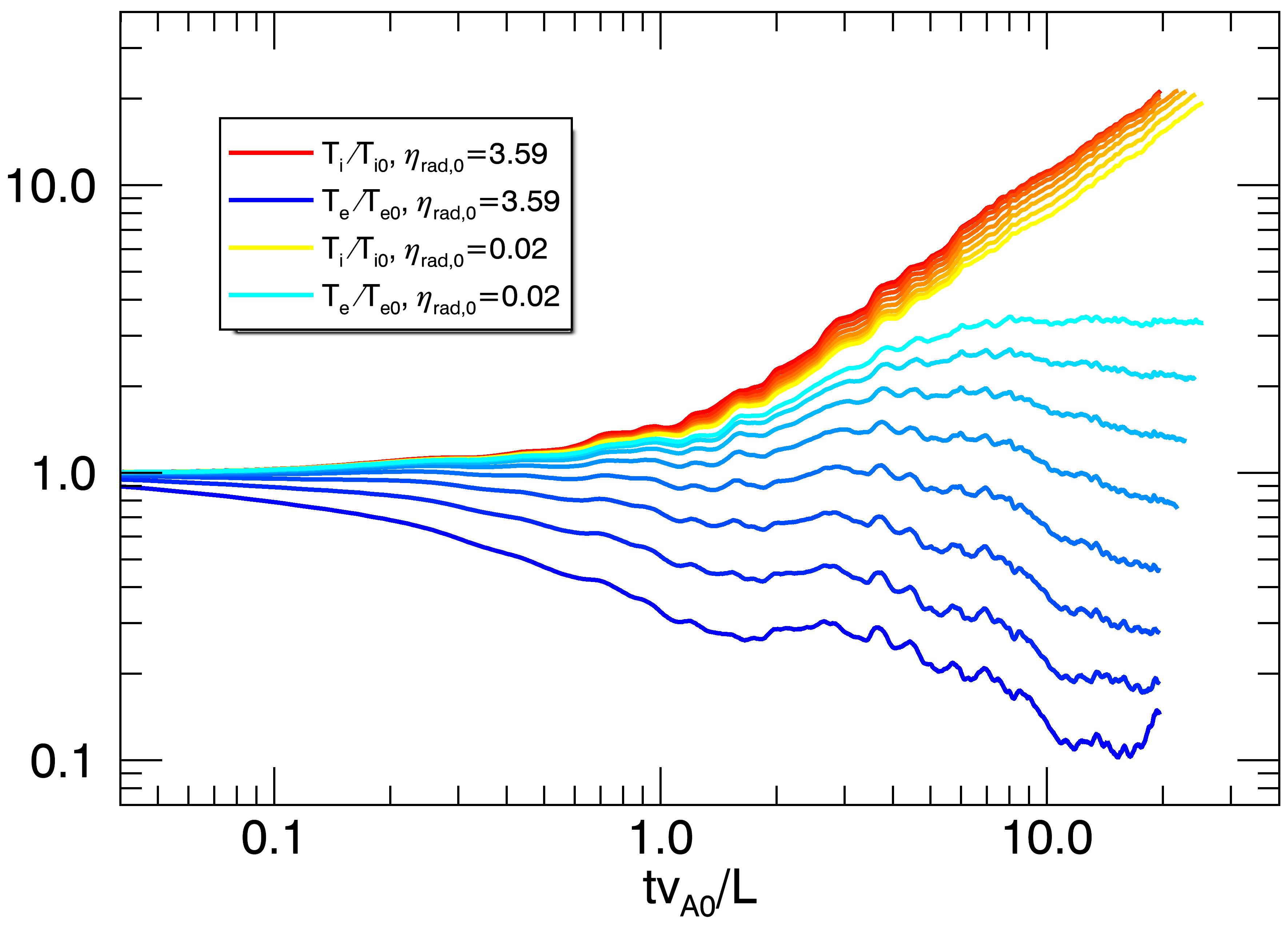}
   \includegraphics[width=\columnwidth]{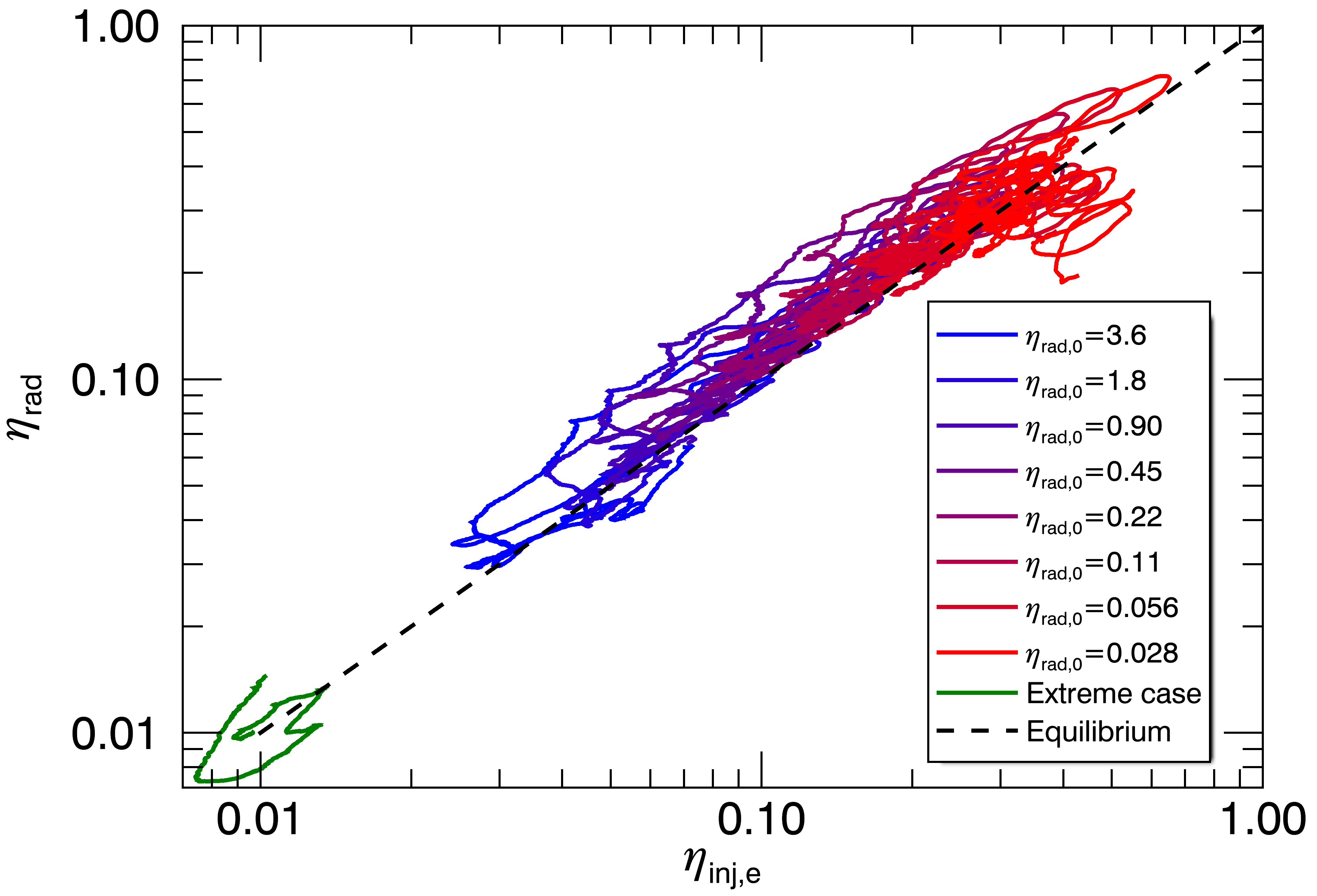}
 \includegraphics[width=\columnwidth]{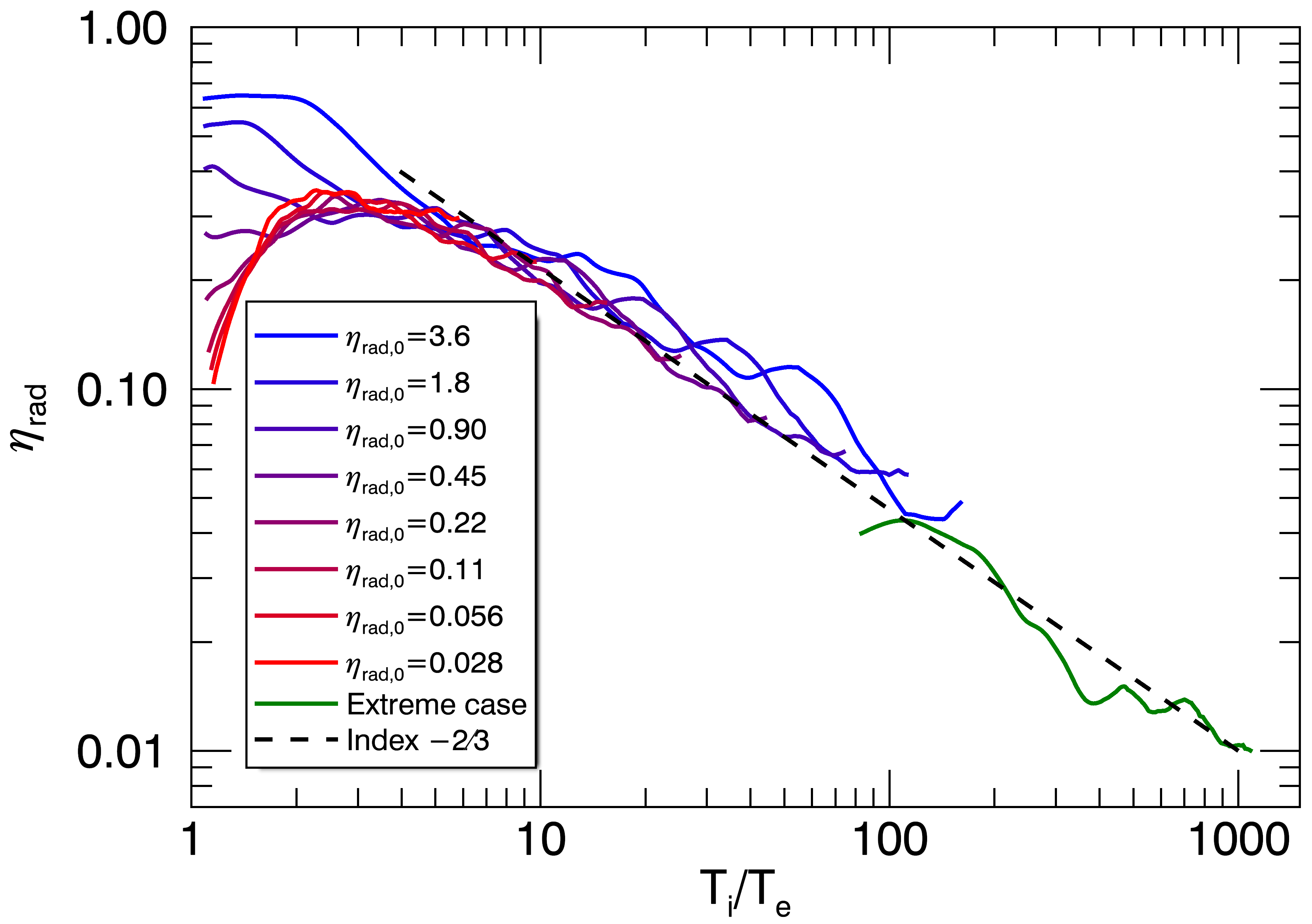}
   \centering
   \caption{\label{fig:etarad} Top: evolution of $T_i/T_{i0}$ (red to yellow) and $T_e/T_{e0}$ (blue to cyan) for varying strength of radiative cooling $\eta_{{\rm rad},0}$ (in order of hue, with bounding cases indicated in the legend). Middle: radiative efficiency $\eta_{\rm rad}$ versus turbulent electron heating efficiency $\eta_{{\rm inj}, e}$, for the same simulations (red-blue) and for the extreme case (green). Bottom: evolution of $\eta_{\rm rad}$ versus $T_i/T_e$, for the same cases, along with a $(T_i/T_e)^{-2/3}$ scaling (dashed line).}
 \end{figure}
 
We use the parameter scan in $\eta_{{\rm rad},0}$ to determine how the temperatures depend on the radiative cooling. As shown in the top panel of Fig.~\ref{fig:etarad}, the evolution of $T_i$ is nearly independent of $\eta_{{\rm rad},0}$, indicating that the ions decouple energetically from the electrons. Meanwhile, $T_e$ quickly adjusts to a value that depends on $\eta_{{\rm rad},0}$, and then slowly decreases at late times. This evolution continues until $\rho_i$ grows to $L/2\pi$. 

We now demonstrate that $T_e$ is determined by a balance between the instantaneous turbulent electron heating and radiative cooling, and this balance slowly shifts to lower temperatures as the electron-to-ion heating ratio decreases with increasing $T_i/T_e$. We compare the radiative efficiency $\eta_{\rm rad} = \dot{\epsilon}_{\rm rad}/\dot{\epsilon}_{\rm inj}$ to the turbulent electron heating efficiency $\eta_{{\rm inj},e} = Q_e/\dot{\epsilon}_{\rm inj}$, which are computed from the instantaneous radiative cooling rate $\dot{\epsilon}_{\rm rad}$, turbulent electron heating rate $Q_e$ (from integrating $\boldsymbol{E} \cdot \boldsymbol{J}_e$ across the domain, where $\boldsymbol{J}_e$ is the electron current density), and external energy injection rate $\dot{\epsilon}_{\rm inj}$. In the middle panel of Fig.~\ref{fig:etarad}, we show the evolution of $\eta_{\rm rad}$ versus $\eta_{{\rm inj},e}$ from $t v_{{\rm A}0}/L = 4$ until the time when $\rho_i = L/2\pi$ for each case in the parameter scan. We also show the extreme case (green line), which reaches $\eta_{\rm rad} \sim 0.01$. We find that $\eta_{\rm rad} \gtrsim \eta_{{\rm inj},e}$, indicating that the radiative cooling and turbulent electron heating are nearly in balance, but radiative cooling is consistently stronger, leading to net cooling at late times.

To uncover the underlying scalings, we show the evolution of $\eta_{\rm rad}$ versus $T_i/T_e$ for the same cases in the bottom panel of Fig.~\ref{fig:etarad}. After a transient establishes $\eta_{\rm rad} \sim 0.3$, the subsequent evolution can be fit by $\eta_{\rm rad} \sim (T_i/T_e)^{-2/3}$. This scaling matches the electron-to-ion heating ratio measured in simulations of non-radiative plasma turbulence, $Q_e/Q_i \sim (\rho_e/\rho_i)^{2/3}$ \citep{zhdankin_etal_2019}. We thus propose that, for larger systems and longer durations, $\eta_{\rm rad}$ (and $T_e$) will continue to decrease as $T_i/T_e$ increases, in accordance with this scaling. Therefore, there is no equilibrium electron temperature.

\newpage

\subsection{Nonthermal particle acceleration}

\begin{figure}
\includegraphics[width=\columnwidth]{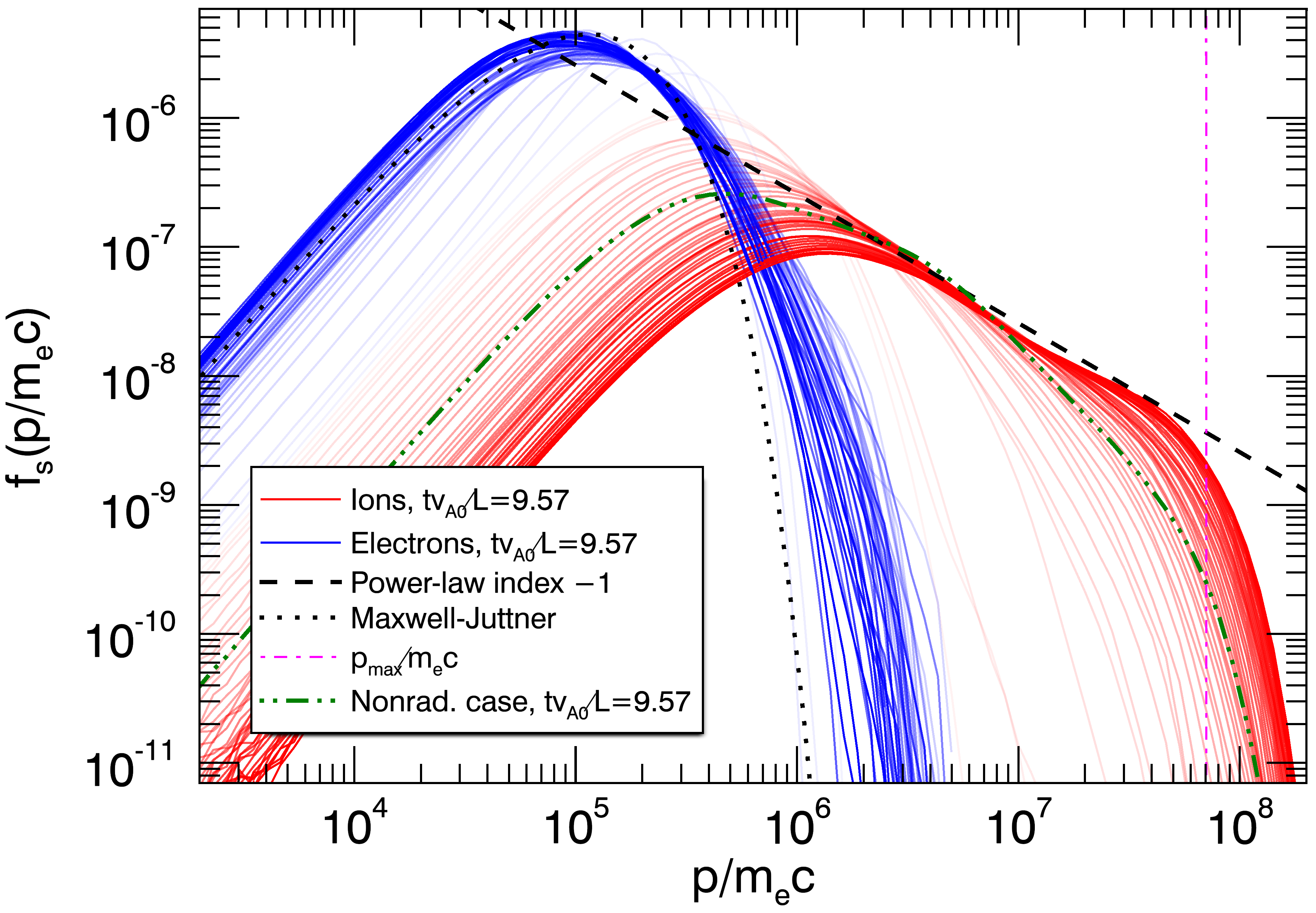}
   \centering
   \caption{\label{fig:dist} Energy distributions for electrons (blue) and ions (red) at varying times, with earlier times indicated by higher transparency. Also shown are the system-size-limited energy $p_{\rm max}$ (magenta, dash-dotted), a Maxwell--J\"{u}ttner distribution fit to the electrons (black, dotted), a power-law with index $-1$ (black, dashed), and the ion distribution from an otherwise identical non-radiative simulation (green, dash-dotted).}
 \end{figure}

We now delve into the kinetic aspects of the plasma. We start with the particle energy distributions, which are identical to (direction-integrated) momentum distributions because particles are ultra-relativistic, $E \approx p c$. In Fig.~\ref{fig:dist}, we show the evolution of the electron energy distribution $f_e(p)$ and ion energy distribution $f_i(p)$. Ions undergo efficient nonthermal particle acceleration, attaining a broad distribution that extends to the system-size-limited momentum, $p_{\rm max} \equiv L e B_0/2 c$ (where the particle gyro-orbit spans the domain). The nonthermal tail is approximately a power law, whose index reaches $-\alpha_i \equiv \partial \log{f_i} / \partial \log{p} \approx -1.4$ rapidly and then becomes progressively shallower as particles accumulate near $p_{\rm max}$. At the latest times, $\alpha_i \lesssim 1$, indicating that the majority of ions are nonthermal. This power law is harder than that obtained from the otherwise identical simulation without radiative cooling (green line in Fig.~\ref{fig:dist}), which has $\alpha_i \approx 2$. Hence, energy is more efficiently channeled into nonthermal ions when an electron-ion scale separation is induced by radiative cooling.

Previous studies have suggested that gyroresonance with MHD-scale fluctuations is the primary acceleration mechanism in relativistic turbulence \citep{zhdankin_etal_2018, comisso_sironi_2019, wong_etal_2020}. From tracked ions, we find that the perpendicular electric field, $\boldsymbol{E}_\perp \equiv \boldsymbol{E} - \boldsymbol{E} \cdot \hat{\boldsymbol{B}} \hat{\boldsymbol{B}}$ (where $\hat{\boldsymbol B}$ is the direction of $\boldsymbol{B}$), accounts for over $98\%$ of the overall ion energy gain, consistent with this picture.

Electrons acquire a quasi-thermal distribution (blue lines in Fig.~\ref{fig:dist}). The bulk is well fit by a Maxwell--J\"{u}ttner distribution, which can be explained by a competition between diffusive acceleration (described by a momentum diffusion coefficient scaling ${\propto}p^2$) and IC radiative cooling \citep[][]{zhdankin_etal_2020}. Aside from the thermal bulk, there is a steep nonthermal tail that spans a factor of a few in energy. From tracked electrons, we find that only ${\sim}60\%$ of the electron energy gain is from $\boldsymbol{E}_\perp$.

\begin{figure}
   \includegraphics[width=\columnwidth]{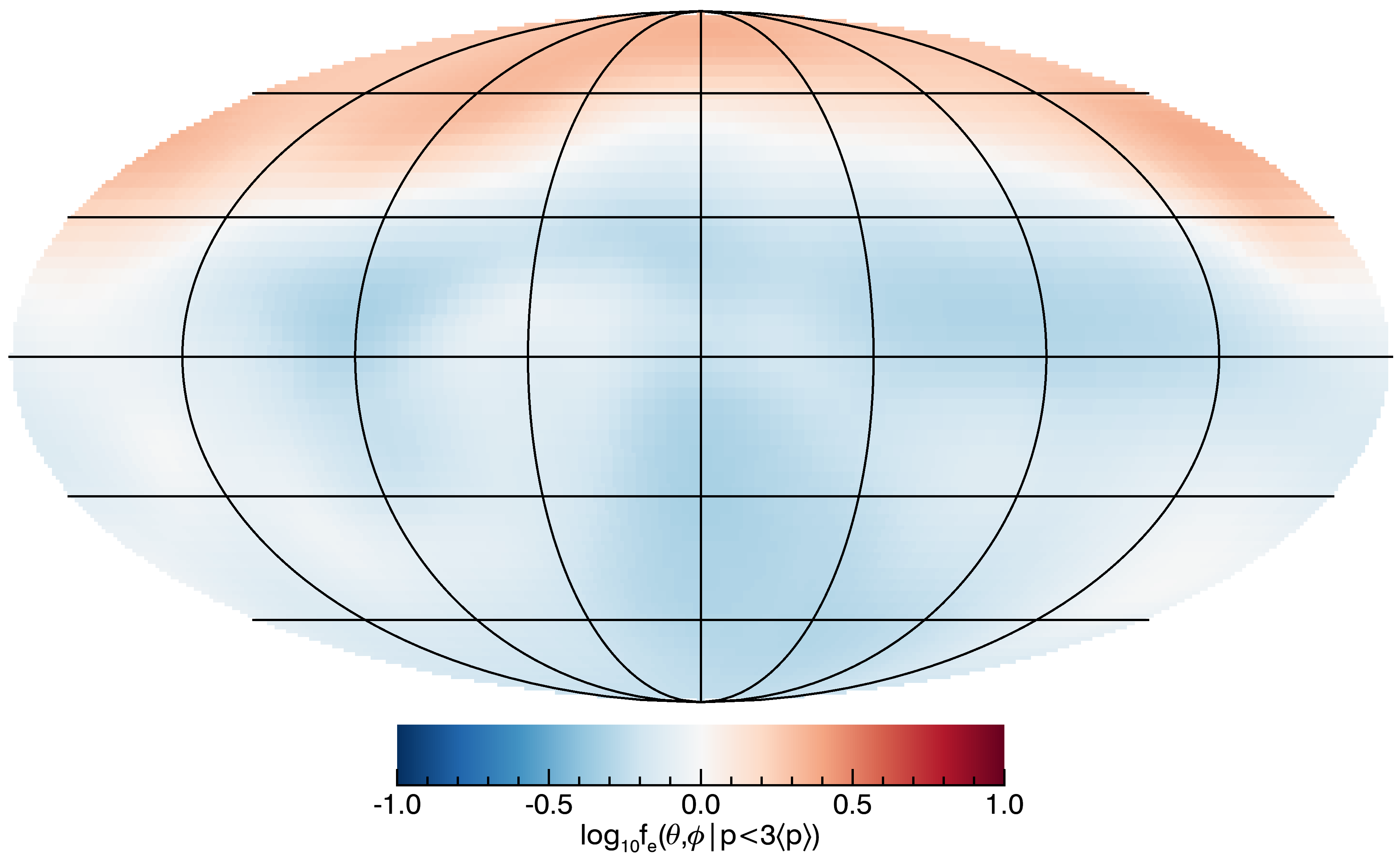}
    \includegraphics[width=\columnwidth]{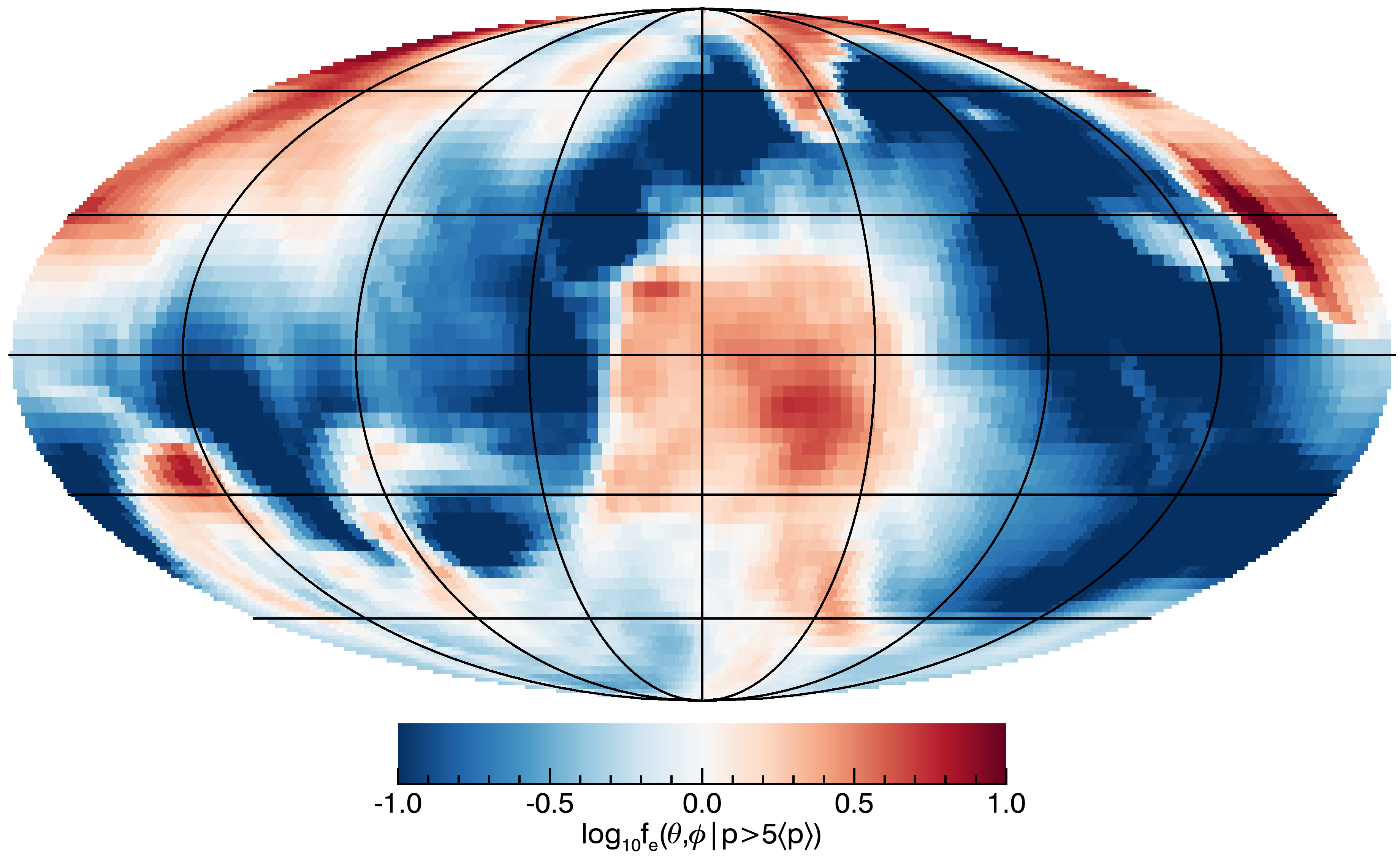}
   \centering
   \caption{\label{fig:dist_ang} Anisotropy of momentum distribution for moderate-energy electrons ($p < 3 \langle p \rangle$; top) and for high-energy electrons ($p > 5 \langle p \rangle$; bottom). The distributions are shown logarithmically and normalized to the direction-averaged value.}
 \end{figure}
 
 \subsection{Intermittent beaming}

Finally, we examine the anisotropy of the momentum distribution. We focus on the electrons, motivated by the fact that IC radiation is emitted in the direction of the relativistic electron motion, and thus electron beams are correlated with observable photon beams. We define the reduced momentum anisotropy distribution, $f_e(\theta,\phi|p_1 < p < p_2)$, as the three-dimensional momentum distribution, $f_e(p,\theta,\phi) = p^2 f_e(\boldsymbol{p})$, integrated across the energy range $p_1 < p < p_2$. Here, $\theta$ is the polar angle with respect to $\boldsymbol{B}_0$ and $\phi$ is the azimuthal angle.

In Fig.~\ref{fig:dist_ang}, we show $f_e(\theta,\phi|p_1 < p < p_2)$ from the fiducial simulation at $t v_{{\rm A}0}/L = 6.4$ for moderate-energy electrons ($p_1 = 0, p_2 = 3 \langle p \rangle$) and for high-energy electrons ($p_1 = 5 \langle p \rangle, p_2 = \infty$), where $\langle p \rangle$ is the mean electron momentum. Following prior studies of radiative magnetic reconnection \citep[][]{cerutti_etal_2013,kagan_etal_2016,mehlhaff_etal_arxiv}, we employ a Mollweide projection to visualize the $(\theta, \phi)$ surface, with the north (south) pole indicating the direction parallel (anti-parallel) to~$\boldsymbol{B}_0$.  The distribution for moderate-energy electrons is nearly isotropic, with a slight random asymmetry due to large-scale flows or currents. The high-energy electrons, by comparison, exhibit substantial small-scale anisotropy, with spikes that are a factor of ${\sim}10$ times the angle average. The pattern of this anisotropy varies on timescales much shorter than $L/v_{{\rm A}0}$. Turbulence in this regime thus produces intermittent beams of electrons, much like in radiative pair plasma turbulence \citep{zhdankin_etal_2020}. Ions also undergo intermittent beaming, but on longer timescales than electrons.

\section{Conclusions}

We analyzed PIC simulations of driven turbulence in collisionless, relativistic, electron-ion plasmas with external IC radiative cooling acting on the electrons. We found no evidence of a collisionless electron-ion thermal coupling mechanism strong enough to maintain $T_i/T_e$ in a steady state. Instead, $T_i/T_e$ is unconstrained, limited only by time and system size, with simulations achieving $T_i/T_e \sim 10^3$. The ions heat up and the electrons cool down; as a consequence, the radiative efficiency $\eta_{\rm rad}$ becomes very low as ions absorb an increasing fraction of the dissipated energy, with simulations achieving $\eta_{\rm rad} \sim 0.01$. The scaling of $\eta_{\rm rad}$ is consistent with our previously proposed empirical formula for the electron-to-ion heating ratio, $\eta_{\rm rad} \sim Q_e/Q_i \sim (\rho_e/\rho_i)^{2/3} \sim (T_e/T_i)^{2/3}$ \citep[see][]{zhdankin_etal_2019}. Electrons acquire a quasi-thermal energy distribution, while ions attain an extended nonthermal distribution with a very hard tail, indicating that turbulence in this regime may be an efficient cosmic-ray accelerator. This ion distribution is significantly harder than in equivalent non-radiative pair plasma simulations \citep[e.g.,][]{zhdankin_etal_2017}. Finally, there is significant intermittent anisotropy in the momentum distribution for high-energy electrons, indicating that turbulence produces electron beams that may manifest as rapid flares.

Future studies will further investigate the nonthermal phenomena highlighted in this Letter. Careful parameter scans are needed to test and develop analytic theories for the ion-to-electron heating ratio, which ultimately determines $\eta_{\rm rad}$. It is also important to access other parameter regimes (e.g., sub-relativistic temperatures and weaker cooling) and incorporate additional radiation channels (i.e., synchrotron).

Our results confirm that the extreme temperature ratios required by some models of radiatively inefficient accretion flows are not unreasonable \citep{shapiro_lightman_eardley_1976, ichimaru_1977, rees_etal_1982, narayan_yi_1995}, although one must carefully take into account the timescales required for turbulence to establish such temperature ratios. Understanding the evolution of $T_i/T_e$ and thus $\eta_{\rm rad}$ in quantitative detail is essential for interpreting emission from the accretion flows in the Galactic Center (around Sgr A*) and in M87, recently detected by the Event Horizon Telescope \citep{eht_2019}. Intermittent electron beams produced by turbulence may be a candidate for explaining rapid X-ray flares in the Galactic Center \citep{baganoff_etal_2001, porquet_2003, eckart_etal_2009}, but more work is required to characterize their statistical properties and reconcile their narrow energy extent with observed broadband spectra. Our study is motivated by accretion flows due to their long-suspected two-temperature nature, but the parameter regime in our simulations may be more directly applicable to giant radio lobes \citep[see, e.g.,][]{erlund_etal_2008, colafrancesco_etal_2011} and relativistic jets from active galactic nuclei.

\acknowledgments

The authors thank G. Werner and M. Begelman for discussions relevant to this project in its early stages. VZ acknowledges support for this work from NASA through the NASA Hubble Fellowship grant \#HST-HF2-51426.001-A awarded by the Space Telescope Science Institute, which is operated by the Association of Universities for Research in Astronomy, Inc., for NASA, under contract NAS5-26555. DU acknowledges support from NASA ATP grants NNX17AK57G and 80NSSC20K0545 and from NSF grant AST-1806084. MK acknowledges support from NSF grant AST-1715277 and from an Alfred P.~Sloan Fellowship in Physics. An award of computer time was provided by the Innovative and Novel Computational Impact on Theory and Experiment (INCITE) program. This research used resources of the Argonne Leadership Computing Facility, which is a DOE Office of Science User Facility supported under Contract DE-AC02-06CH11357. This work also used the Extreme Science and Engineering Discovery Environment (XSEDE), which is supported by National Science Foundation grant number ACI-1548562. This work used the XSEDE supercomputer Stampede2 at the Texas Advanced Computer Center (TACC) through allocation TG-PHY160032 \citep{xsede}.


\software{Zeltron \citep{cerutti_etal_2013}}

\end{document}